\documentstyle[aps,psfig]{revtex}
\newcommand{\be}{\begin{equation}}
\newcommand{\ee}{\end{equation}}
\newcommand{\bea}{\begin{eqnarray}}
\newcommand{\eea}{\end{eqnarray}}
\newcommand{\non}{\nonumber}
\newcommand{\MeV}{\,{\rm MeV}}
\newcommand{\fm}{{\,\rm fm}}
\newcommand{\re}{{\rm Re}\,}
\newcommand{\im}{{\rm Im}\,}
\newcommand{\dn}{{\rm d}}
\newcommand{\bm}[1]{\mbox{\boldmath$#1$}}
\newcommand{\cauch}{{\cal P}\hspace{-0.2cm}}

\begin{document}
\draft
\twocolumn[\hsize\textwidth\columnwidth\hsize
           \csname @twocolumnfalse\endcsname
\title{Precritical pair fluctuations and formation of a pseudogap\\
in low-density nuclear matter}
\author{A. Schnell and G. R\"opke}
\address{University of Rostock, FB Physik, Universit\"atsplatz 1,
18051 Rostock, Germany}
\author{P. Schuck}
\address{Institut des Sciences Nucl\'{e}aires,
Universit\'{e} Joseph Fourier, CNRS-IN2P3\\
53, Avenue des Martyrs, F-38026 Grenoble Cedex, France
}
\date{\today}
\maketitle
\begin{abstract}
Low-density nuclear matter at finite temperature is considered representing the strong coupling situation of a highly correlated fermion system. One-particle self-energies and the density of states in the vicinity of the pairing transition point are presented. Within the Green function approach model calculations are performed using the thermodynamic $T$ matrix in ladder approximation. As a main result the formation of a pseudogap in the level density near the critical temperature of the pairing phase transition has been found. It is shown that compared to mean field, the picture of the onset of the pairing transition is essentially changed if correlations are taken into account.

\end{abstract}
]
\newpage
Among strongly coupled fermion systems which show a transition to a superfluid or superconducting state \cite{BEC} such as, e.g., the electron gas in metals, condensed ${\rm He}^3$, or electron-hole pairs in excited semi-conductors, low-density nuclear or neutron matter turns out to be a particularly strongly correlated system. According to BCS theory \cite{BCS} the ratio $\Delta/\epsilon_F$ ($\Delta$ being the BCS gap, $\epsilon_F$ the Fermi energy) is of the order $10^{-3}$ in ordinary metals whereas it is $0.1$ in high $T_c$ superconductors. In finite nuclei the ratio is about $0.03$ and thus much closer to the scenario of high $T_c$ superconductors than to normal superconductivity. Indeed, the coherence length $\xi=v_F\hbar/\Delta$ for nucleons in the spin singlet ($S=0$), isospin triplet ($T=1$) channel becomes of the same order as the internucleon distance at densities $\rho\approx\rho_0/10\dots\rho_0/5$ with $\rho_0=0.17\fm^{-3}$ the nuclear matter saturation density. In the $S=1$, $T=0$ neutron-proton channel (the deuteron), where the interaction is more attractive, the correlations are even stronger.

Referring to this, a very fundamental quantity is the two-particle spectral function which is governed by such correlations. At the same time, the two-particle spectral function is an important ingredient to determine single-particle properties. In particular, in high-$T_c$ materials such as underdoped cuprates, strong correlations are present in the electron-hole system of the normal (nonsuperconducting) phase leading to the formation of the so-called pseudogap in the charge and spin excitation spectra. It is known from experiments and from theoretical studies that the pseudogap occurs below a certain temperature $T^*$ having similar characteristics as the superconducting gap which grows continuously out of the pseudogap \cite{HTC}. It appears that the pseudogap is a precursor of the superconducting gap. 

In this letter we focus on low-density nuclear matter at finite temperatures. Our main point of interest is the isospin singlet pairing \cite{3S1,ParisGap} being of relevance mostly in symmetric nuclear matter \cite{asymm} which will be studied here. It should be pointed out that low-density nuclear or neutron matter not only exists during stellar collapse or in the outer sphere of neutron stars but is an important prerequisite to study pairing effects via the rather widespread and well accepted local density approximation also in the surface layer of real nuclei. There, evidence of strong correlations is given by electron scattering experiments off nuclei \cite{escatt} where the nucleon spectral function and momentum distributions are determined. In such highly correlated systems the mean-field BCS theory is not the appropriate tool for the description of pairing effects.

It is therefore of great importance to include correlations in the theoretical treatment of such systems. We use the $T$ matrix approach within the Green function theory \cite{KB62,FW71}. This is a very powerful tool to describe bound states (deuterons) in the normal state and the pairing transition (formation of Cooper pairs) \cite{asymm,SRS90,ARS94,ARSS95,spectral}.

The single-particle propagator obeys the Dyson equation \cite{KB62,FW71}
\be\label{dyson}
G_1(\vec{p}\vec{p}',z) = \frac{\delta_{\vec{p}\vec{p}'}}
{z-p^2/(2m)-\Sigma(p,z)}\,,
\ee
where the self-energy $\Sigma$ contains the influence of the medium on the propagation of a single particle which is systematically accounted for in the single-particle spectral function $A(p,\omega)$, based on which the level density is given by
\be\label{levelspec}
N(\omega)=\sum_pA(p,\omega)\,.
\ee
In the following we are interested in energies around the Fermi level and therefore we will neglect the imaginary part of $\Sigma$, i.e. the spectral function is approximated by a $\delta$-function at the quasi-particle energy $\epsilon(p)$ given by
\be\label{eps}
\epsilon(p) = \frac{p^2}{2m}+U(p)\,;\qquad
U(p) = \re\Sigma(p,\epsilon(p))\,.
\ee
Then the level density can be calculated via
\be\label{Def}
N(\omega) = \sum_{p}\delta(\omega-\epsilon(p))\,,
\ee
which is identical to (\ref{levelspec}), see e.g.\ \cite{Dickhoff}. The calculation of (\ref{Def}) yields the final expression for the density of states
\be\label{leveldens}
N(\omega) = \left[\frac{1}{p^2}
\frac{\partial\epsilon(p)}{\partial p}\right]^{-1}
\,;\qquad\omega = \epsilon(p)\,.
\ee
It is completely determined by the real part of the on-shell self-energy and related to the effective mass $m^*(p) = 2\,N(\omega)/p$. The consistent inclusion of $\im\Sigma$ is in progress.

Within a cluster decomposition \cite{cluster} the single-particle self-energy can be calculated from the two-particle $T$ matrix $T_2(12,1'2',z)$ taking only two-particle correlations into account. One obtains for the real part \cite{SRS90}
\begin{mathletters}
\label{resigall}
\bea
\re\Sigma(p_1,\omega) & = & \re\Sigma^{(1)}(p_1,\omega)+
\re\Sigma^{(2)}(p_1,\omega)\,,\label{resigsum}\\
\re\Sigma^{(1)}(p_1,\omega) & = &
\sum_2f(2)\re T_2(12,12,\omega+\epsilon(p_2))\,,\label{resiga}\\
\re\Sigma^{(2)}(p_1,\omega) & = &\!\!
\sum_2\!\cauch\int\!\!\frac{\dn\omega'}{\pi}\frac{g(\omega')
\im T_2(12,12,\omega'+i0)}{\omega'-\omega-\epsilon(p_2)},
\!\label{resigb}
\eea
\end{mathletters}
with $f(2)=f(\epsilon(p_2))=[\exp\{(\epsilon(p_2)-\mu)/T\}+1]^{-1}$ being the Fermi and $g(E)=[\exp\{(E-2\mu)/T\}-1]^{-1}$ the Bose distribution function. The two-particle ladder $T$ matrix is given by the Galitskii-Feynman equation \cite{FW71}
\bea
T_2(12,34,z) & = & V(12,34)+\sum_{1'2'}V(12,1'2')\nonumber\\
&&\times G_2^0(1'2',z)\,T_2(1'2',34,z)\,.
\label{BGE}\eea
The uncorrelated two-particle Green function $G_2^0$ is given as a product of two single-particle propagators via the one-particle spectral function. Using the quasi-particle approximation as mentioned above one obtains
\be\label{G20}
G^0_2(12,z)=\frac{1-f(1)-f(2)}
{z-\epsilon(p_1)-\epsilon(p_2)}\,.
\ee
Equations (\ref{eps}) and (\ref{resigall}) together with (\ref{BGE}) and (\ref{G20}) form a self-consistent set of equations which is solved iteratively.
The usual first-order Br\"uckner-Hartree-Fock approximation \cite{BHF} can be recovered replacing in (\ref{G20}) the expression $1-f(1)-f(2)$ by $(1-f(1))(1-f(2))$, i.e., neglecting the ``backward going'' part. As we will demonstrate later the inclusion of the hole-hole sector is, however, essential to account for precritical pair fluctuations.

For an exploratory calculation we use a separable representation of the nucleon-nucleon potential according to Yamaguchi \cite{YamaPot}
\be\label{seppot}
V_\alpha(p,p') = -\frac{\lambda_\alpha}{\Omega}v(p)v(p')\,;
\qquad v(p) = \frac{1}{p^2+\beta^2}\,,
\ee
with the relative momenta $p$ and $p'$, $\alpha$ denotes the interaction channel, $\lambda_\alpha$ the coupling strength of the respective channel, and $\Omega$ the normalization volume. The parameters $\beta=1.4488\fm^{-1}$ and $\lambda_\alpha=151.7\MeV\fm^3$ for the $^1{\rm S}_0$ channel and $\lambda_\alpha=216.1\MeV\fm^3$ for the $^3{\rm S}_1$ channel are chosen to reproduce the low-energy scattering phase shifts as well as the deuteron binding energy. This potential yields somewhat too strong pairing properties as compared to those calculated, e.g.\ with the Paris potential \cite{ParisGap}, but this will not invalidate the important conclusions which we will draw from our study below. Using the potential (\ref{seppot}) a solution of (\ref{BGE}) is particularly simple. Transforming to relative $\vec{p} = (\vec{p_1}-\vec{p_2})/2$ and total momentum $\vec{P} = \vec{p_1}+\vec{p_2}$ one obtains
\bea
T_\alpha(pp'P,z) & = & \frac{V_\alpha(p,p')}
{1-J_{\alpha}(P,\mu,T,z)}\,,\\
J_{\alpha}(\mu,T,P,z) & = & \frac{-\lambda_\alpha}
{2\pi^2}\int\dn p\,p^2v^2(p)G_2^0(p,P,z)\,.
\eea

Since $\re\Sigma^{(2)}$ (\ref{resigb}) is of particular interest we will consider it in more detail. In particular, there is a connection to the two-particle spectral function $A_2$ that is defined as the imaginary part of the full two-particle Green function $G_2$. Instead of $A_2(p_1p_2,p_1'p_2',\omega)$ we will consider the contraction ($\vec{p}_{1,2}=\vec{P}/2\pm\vec{p}\,,\;\vec{p}\,'_{1,2}=\vec{P}/2\pm\vec{p}\,'$)
\be\label{a2}
F_2(P,\omega)=\sum_{pp'}v(p)\im G_2(\vec{p}_1\vec{p}_2\vec{p}\,'_1\vec{p}\,'_2,\omega+i0) v(p')\,,
\ee
which is a function of only the total momentum and the energy. With (\ref{a2}) one can rewrite (\ref{resigb})
\be
\re\Sigma^{(2)}(p_1,\omega) =
\sum_2\cauch\int\frac{\dn\omega'}{\pi}
V^{-1}(12,12)\frac{g(\omega')F_2(P,\omega')}
{\omega'-\omega-\epsilon(p_2)}\,.
\ee
It can be shown that $F_2$ obeys the following energy-weighted sum rule (the analogue to the $f$-sum rule in the particle-hole channel)
\bea
\int\limits_{-\infty}^{+\infty}\frac{\dn\omega}{\pi}
\omega F_2(P,\omega) & = &
\int\frac{\dn^3\bm{p}}{(2\pi)^3}\left(\frac{p^2}{m}+
\frac{P^2}{4m}\right)
Q(p,P)v^2(p)\non\\
&&{}+\left(\int\frac{\dn^3\bm{p}}{(2\pi)^3}v^2(p)Q(p,P)\right)^2
\label{sumrule}
\eea
($Q$ denotes the angle averaged Pauli blocking operator).

In the following we are going to present results of a calculation for symmetric nuclear matter at the density $\rho=\rho_0/3$. Using the self-consistent procedure described above one obtains from the Thouless criterion of the $T_2$ matrix \cite{SRS90} a critical temperature $T_c=3.6\,\MeV$ in the $^3{\rm S}_1$-channel.
\begin{figure}[tb]
\centerline{\psfig{figure=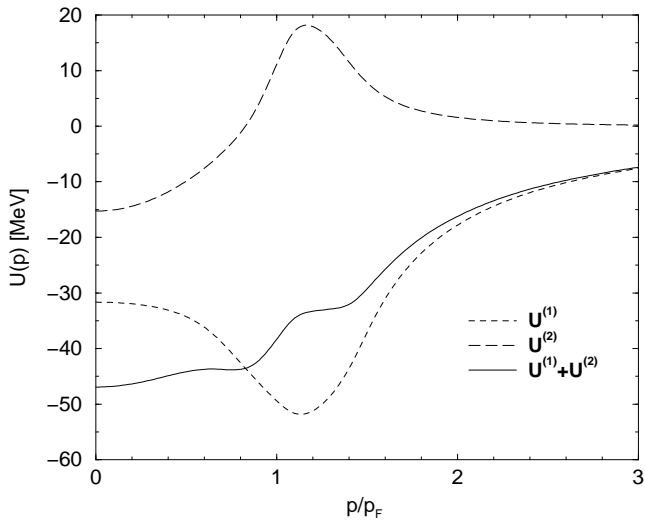,width=8.5cm}}
\caption{Single-particle potential (\ref{eps}) of the nucleon in nuclear matter as a function of the momentum $p$ over the Fermi momentum $p_F=0.92\fm^{-1}$ at the density $\rho=\rho_0/3$ and temperature $T=5\,\MeV$. The contributions $U^{(1)}$ and $U^{(2)}$ are defined according to Eqs.~(\ref{eps}), (\ref{resiga}), and (\ref{resigb}), respectively.}
\label{fig-resig}
\end{figure}
In our example we chose the temperature $T=5\MeV$ being close but above this critical value. In Fig.~\ref{fig-resig} the self-consistent real part of the on-shell self-energy (single-particle potential) $U(p)$ is displayed as a function of the momentum $p$ showing the contributions $U^{(1)}$ and $U^{(2)}$ and their total sum according to Eqs.~(\ref{resiga}), (\ref{resigb}), and (\ref{resigsum}), respectively. The sum $U(p)=U^{(1)}(p)+U^{(2)}(p)$ is a delicate balance of the first and the second contribution to the single-particle potential. Although each contribution shows a smooth behavior the sum of both exhibits two plateaus around the Fermi momentum $p_F$ sandwiching a steeper slope. This effect can be traced back to the pairing instability of the $T_2$ matrix that is very pronounced already close to $T_c$ (see \cite{ARS94,spectral}). A similar result has been found at zero temperature by Vonderfecht et al. \cite{Vonderfecht} who calculated $U(p)$ in the same approach but under consideration of the BCS gap in order to avoid the pairing instability.

Because of the subtlety of the effect it is important to check numerical accuracy. We therefore calculated the real part of the self-energy also via dispersion relation from the imaginary part according to
\be\label{disp}
\re\Sigma(p_1,\epsilon(p_1)) = \Sigma^{\rm HF}(p_1)\!+
\!\!\int\!\frac{\dn\omega}{\pi}
\frac{\im\Sigma(p_1,\omega+i0)}{\omega-\epsilon(p_1)}\,.
\ee
$\Sigma^{\rm HF}$ denotes the Hartree-Fock shift and $\im\Sigma$ is given by
\bea
\im\Sigma(p_1,\omega) & = & \int\frac{\dn^3p_2}{4\pi}
\im T_2(12,12,\omega+\epsilon(p_2)+i0)\non\\
&&\qquad\times{}[f(\epsilon(p_2))+g(\omega+\epsilon(p_2))]\,.
\label{imsig}
\eea
Within the bounds of numerical accuracy the calculation of $\re\Sigma$ according to (\ref{disp}), (\ref{imsig}) yielded equal results compared to the procedure following from (\ref{resigall}).
\begin{figure}[t]
\centerline{\psfig{figure=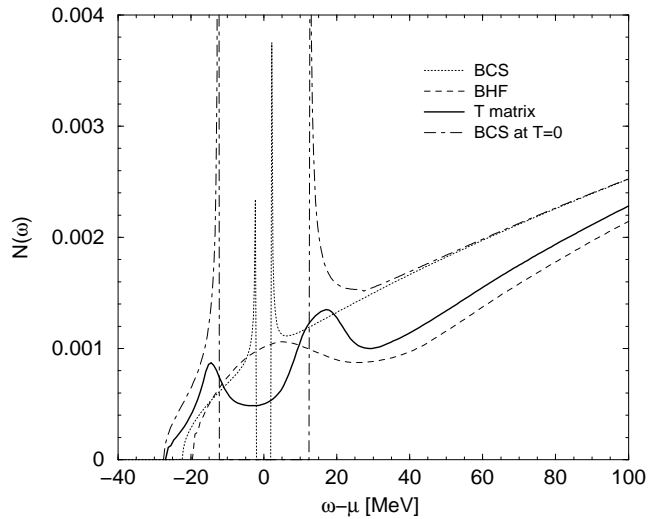,width=8.5cm}}
\caption{Nucleon level density (\protect\ref{leveldens}) as a function of the energy $\omega-\mu$ normalized to the chemical potential $\mu$ given for the density $\rho=\rho_0/3$ and temperature $T=5\MeV$. Results of three different approximations are shown: $T$ matrix approach (solid) and for comparison Br\"uckner-Hartree-Fock (BHF, dashed) and mean-field BCS theory (dotted). The dot-dashed curve is a $T=0$ calculation in BCS approximation.}
\label{fig-leveldens}
\end{figure}

According to (\ref{leveldens}) the behavior of the single-particle potential is closely related to the density of states. In Fig.~\ref{fig-leveldens} $N(\omega)$ is plotted as a function of $\omega-\mu$ for the same values of density and temperature as in Fig.~\ref{fig-resig}. One observes a depletion of the density of states around the chemical potential and the rise of two pronounced peaks. This formation of a so-called pseudogap (see below) in the level density is a precursor signal of the superfluid phase transition expected to occur at $T_c=3.6\MeV$. For comparison, in Fig.~\ref{fig-leveldens} results of other approximations are given. The mean-field BCS approach at $T=5\MeV$ (dotted curve) which is closely below the critical temperature of $T_c^{\rm BCS}=5.08\MeV$  exhibits a tiny gap of twice the BCS gap at Fermi momentum [$\Delta(p_F)=2.1\MeV$] around the chemical potential. The BCS result for zero temperature (dot-dashed) shows a considerably larger gap. There is a striking similarity between the BCS result at zero temperature and the $T$ matrix pseudogap with respect to the location of the two peaks. We also calculated the level density in Br\"uckner-Hartree-Fock (BHF) approximation \cite{BHF} which neglects the hole-hole contribution to the effective interaction which is necessary for the description of pairing. Interestingly, the BHF calculation yields an opposite behavior compared to the $T$ matrix approach which is known as the enhancement of the effective mass at the Fermi surface \cite{BHF}.

In the following we want to discuss the consequences of our calculation in detail. The occurrence of the pseudogap in the density of state is a phenomenon which has recently been discovered experimentally and discussed theoretically in the context of high $T_c$ superconductors \cite{HTC}. Some authors attribute the measured depression in the density of states to pair fluctuations in analogy to the way described here \cite{Randeria}. In the nuclear physics context our finding is novel and certainly opens up a so far unrevealed aspect of the nuclear density of states.

However, the situation in finite nuclei is more complex. One should realize that all the effort over the last decades concerning the nuclear level density parameter rather went into the opposite direction trying to explain why the experimental single-particle level density of finite normal fluid nuclei around the Fermi energy is enhanced rather than diminished with respect to the mean-field result. Indeed, in including long range particle-hole correlations such as surface vibrations, it is very well known \cite{Schuck} that the single-particle potential is flattened in the vicinity of the Fermi energy rather than steepened as here. The strong collectivity of surface vibrations in nuclei is certainly a quite particular feature of the finiteness of the system which makes that finally the particle-hole correlations in nuclei win over the particle-particle ones (otherwise there would be no deformed nuclei). However, the location of the transition point to stable deformation results from a detailed balance between the battle of pair and shape fluctuations. We believe that apart from rare occasions \cite{VinhMau} the study of the influence of pair vibrations on the nuclear level density is very much underrepresented in the nuclear physics literature and further investigations are of great importance for the detailed comprehension of nuclear systems. Indeed, e.g.\ for finite nuclei at $T=0$ the average level spacing takes over the role of temperature. There exists a critical level spacing from where on the nucleus becomes superfluid. Before, there exist precritical pair fluctuations quite analogous to the situation in infinite matter at $T\sim T_c$. On the other hand in extended objects like neutron star matter it is not clear whether particle-hole correlations win over particle-particle correlations at all. The widely believed scenario where the effective nucleon mass in nuclear matter is enhanced at the Fermi surface \cite{BHF} needs a more critical analysis once one allows for pairing correlations. This is a very important issue for future studies.

In short, in this work we have for the first time performed a self-consistent Galitskii-Feynman-Hartree-Fock calculation at finite temperature approaching the superfluid phase transition point $T_c$ from above. In parallel with theoretical work concerning high $T_c$ superconductivity we also found in low-density nuclear matter the formation of a pseudogap in the density of states. This new finding may have not only important consequences for the physics of neutron and collapsing stars but may also effect details of finite nuclei. Further work is needed to settle the subtle competition between pair and density fluctuations in this respect. Our results open up an essential new aspect in the important debate around the nuclear density of states.

We acknowledge useful discussions with M.\ Baldo, W.\ H.\ Dickhoff, U.\ Lombardo, P.\ Nozi\`{e}res, and N.\ Vinh Mau. Preliminary results with respect to the pseudogap in nuclear matter have been shown in \cite{LesHouches}.

\end{document}